 \documentclass[12pt,preprint]{aastex}

\slugcomment{Submitted to ApJ}
\shorttitle{Local Interstellar Magnetic Field}\shortauthors{Frisch}
\newcommand{\commentpcf}[1]{}

\def\muG{$\mu $G}

\def\glong{$\ell$}
\def\glat{$b$}

\def\PAgal{PA$_\mathrm{G}$}

\def\PAecl{PA$_\mathrm{E}$}

\def\NHI{$N$(H$^{\rm o}$)}

\def\HI{H$^{\rm o}$}
\def\DI{D$^{\rm o}$}

\def\HeI{He$^{\rm o}$}

\def\nHI{$n \mathrm{(H^\circ)}$}
\def\nel{$n \mathrm{(e)}$}

\def\Bis{${B_\mathrm{IS}}$}
\newcommand{\Pol}{$P$}
\def\P5{$P_5$}

\def\Bhat{$\hat{k_\mathrm{B}}$}

\def\el{\hbox{$\lambda$}}
\def\eb{\hbox{$\beta$}}

\def\lambdazero{\hbox{$\lambda_\mathrm{0}$}}

\def\kms{\hbox{km s$^{-1}$}}
\def\deeg{\hbox{$^{\rm o}$}}

\def\Lya{\hbox{Ly$\alpha$}}
\def\HeI{He$^{\rm o}$}

\def\cmtwo{cm$^{-2}$}
\def\thetac{$\theta_{\rm C}$}

\def\cc{cm$^{-3}$}

\begin{document}                                                                

\title{ Interstellar Dust and Magnetic Field at the
Heliosphere} \author{Priscilla C. Frisch} \affil{Department of
Astronomy and Astrophysics, University of Chicago, Chicago, IL 60637.}
\email{frisch@oddjob.uchicago.edu}

\begin{abstract}

The very weak polarization of light from nearby stars reaches a
maximum near the ecliptic plane in a position offset by $ \lambda \sim
35^\circ$ from the heliosphere nose direction.  The position angle for
the polarization in this ecliptic-plane peak is the same for near and
distant stars, to within uncertainties, indicating the interstellar
magnetic field direction is relatively constant over several hundred
parsecs in this region. This region is near the center of a magnetic
bubble that has recently been identified by Wolleben.  The magnetic
field directions defined by the polarizations of nearby and distant
stars in the direction of this polarization maximum, and also by the
observed offset between interstellar \HeI\ and \HI\ flowing into the
heliosphere, converge to a local magnetic field direction that is
inclined by $\sim 55^\circ$ with respect to the galactic plane, or
$\sim 65^\circ$ with respect to the ecliptic plane.  
Surprisingly, the geometry of the cosmic microwave
background (CMB) dipole moment shows a strong correlation with
this heliosphere geometry. Several vectors associated with the CMB low-$\ell$
multipole moments are located at the positions of heliosphere
landmarks such as the 3 kHz emissions detected by the Voyager 1 and
Voyager 2 satellites.  Together these results suggest 
that an unrecognized heliospheric
foreground is present in the measured CMB background.  All data are consistent
with the expansion of the S1 subshell of the Loop I magnetic bubble to
the solar location.  The magnetic field in this shell then determines the
field direction at the Sun, helps shape the heliosphere, and generates
an additional unrecognized CMB foreground.

\end{abstract}

\section{Introduction} \label{sec:intro}

Interstellar dust grains location are a potential
diagnostic of the interaction between the solar wind and the interstellar
medium.  The relative velocity between the Sun and the surrounding
interstellar cloud is 26.3 \kms, and neutral interstellar
gas and large dust grains flow through the 
raindrop-shaped solar wind bubble known as the heliosphere.  Interstellar ions and the
small interstellar dust grains that dominate interstellar polarization
are trapped in the
interstellar magnetic field that is deflected around the heliosphere.

The fundamental interstellar properties that dominate the 
shape of the heliosphere are the direction and strength of the 
magnetic field \Bis, the ionization level, and the thermal and ram 
pressure of gas \citep[e.g.][]{Holzer:1989}.
The upstream directions of interstellar \HI\ and \HeI\ inside the 
heliosphere differ by 5 \deeg, serving as a diagnostic of
the interstellar magnetic field at the Sun 
\citep{WellerMeier:1974,Quemeraisetal:2000,Lallementetal:2005}.  
A second diagnostic of
\Bis\ at the heliosphere is the polarization of light from nearby
stars in the galactic center hemisphere, discovered by
\citep[][T82]{Tinbergen:1982} and now interpreted as partly caused by
interstellar dust grains trapped in the outer heliospheath regions
\citep[][Paper I]{Frisch:2005L}.  The polarization maximum occurs
for stars near the ecliptic plane and located at ecliptic longitudes of
about $\sim +35^\circ$ from the upstream
directions given by interstellar dust and \HeI.

Three independent sets of data indicate that the heliosphere
shape is not symmetric around the upwind direction of interstellar
material (ISM) flowing into the heliosphere.  Voyager 1 crossed the
solar wind termination shock (TS) at 94 AU in December 2004, and
Voyager 2 crossed the TS at $\sim  84$ AU in August 2007, indicating the TS
is closer by $\sim 10$ AU in the southern ecliptic compared to northern
ecliptic regions
\citep{Stone:2007,Stoneetal:2005,Burlagaetal:2005,Deckeretal:2005}.
These asymmetries result from an interstellar magnetic field inclined
to the velocity of inflowing ISM \citep{Pogorelovetal:2004,Opheretal:2007}.
Observations of low frequency radio emissions (1.8--3.6 kHz) by the
plasma wave instruments on Voyager 1 and 2 during 1992-1994 showed
that these events arise $\sim 130$ AU from the Sun at the edge of the
heliosphere, and that the events are consistent with an asymmetric
heliosphere \citep[][Appendix
\ref{app:khz}]{KurthGurnett:2003,GurnettKurthetal:2006}.  The third
data set consists of measurements of $\sim 1$ keV energetic neutral atoms
(ENAs) near the heliosphere nose, that have an origin 
offset to positive ecliptic longitudes compared to the heliosphere
nose direction of \el$\sim 10^\circ - 40^\circ$
\citep{Collieretal:2004,Wurzetal:2004}.  ENAs are formed beyond the termination shock
by charge exchange between interstellar \HI\ and the solar wind; 
the ENAs show the bulges in the solar wind such as the
heliotail and the neutral current sheet wrapping around the
heliosphere nose \citep{Heerikhuisenetal:2007}.

The asymmetry of the outer heliosphere may vary with time
since the solar magnetic activity cycle causes the
solar and interstellar magnetic field lines to alternate between parallel and
antiparallel modes
\cite[e.g.][]{NerneySuess:1995,Macek:1990,WashimiTanaka:1996,Zurbuchen:2007}.
The Tinbergen polarization data and 3 kHz data were collected during
north-pole-positive polarities, with field lines emerging at the solar
north pole.  The Voyager satellites crossing of the termination shock
occurred during solar cycle phases when the south pole had a
positive polarity.

The structure of this paper is as follows.
The polarization data that yield the local magnetic field
direction are discussed in \S \ref{sec:data}, with more
details in Appendix \ref{app:pa}..  The polarization data used in Paper I are
augmented with additional data acquired during the 1970's.
The question as to whether
this weak polarization originates at the Sun or in nearby ISM in the upwind
direction is discussed in \S \ref{sec:origin}. 
The magnetic field direction of nearby and distant stars in
the ecliptic nose region for near and far stars are found to be similar
(\S \ref{sec:blocal}), and yield a magnetic field direction at the Sun
that is independent of whether the polarization is formed at the heliosphere or in the nearest upwind
ISM (\S \ref{sec:bfield}).  Several characteristics of 
the low-$\ell$ components of the cosmic
microwave background that are related to the heliosphere are discussed
in \S \ref{sec:cmb}, including the area vectors that occur
in the direction of the 3 kHz emissions,
the relation of the hot and cold poles of the CMB Doppler dipole
to the heliosphere, and the distribution of nearby ISM. 
The influence on the local magnetic field direction of the S1 magnetic bubble 
is discussed in \S \ref{sec:loop}.
Conclusions are presented in \S
\ref{sec:conclusions}.  Appendix \ref{app:pa} presents additional
details about the optical polarization data, and discusses the
coincidental correlation between the position angles of distant stars
and ecliptic latitude.  Appendix \ref{app:khz} briefly discusses primary
versus alternate locations found for the 3 kHz emissions observed by
the Voyager satellites.

The conclusions in the following discussions are sensitive to the physical 
properties of the ISM surrounding the Sun.  The ISM at the solar location has a temperature of 6300$\pm$340 K, 
a heliocentric velocity of --26.3 \kms, and an upstream direction near the ecliptic
plane at \el=254.7$\pm$0.5\deeg, \eb=5.1$\pm$0.2\deeg\ (corresponding
to \glong=3.5\deeg, \glat=15.2\deeg), based on Ulysses interstellar 
\HeI\ data \citep[][W04]{Moebiusetal:2004,Witte:2004}.
The ISM flow through the
heliosphere contains dust grains, which are observed at the same
velocity and flow direction as the gas
\citep[][]{Gruenetal:1993,Baguhletal:1996,Gruenetal:2005,Frischetal:1999}.
The dust-to-gas mass ratio for the ISM flowing into the
heliosphere is $<120$, compared to $>180$ for the LIC (SF07).
Models that evaluate the ionization gradient in the surrounding ISM
yield interstellar \HI\ densities at the Sun of \nHI=0.19 \cc, and
electron densities \nel$=0.07 \pm 0.01$ \cc\ \citep[e.g. Model 26
in][SF07]{SlavinFrisch:2007}.  The plasma oscillation frequency of ISM
at the Sun is then $\sim 2.4$ kHz, but will rise to higher values in
the interstellar plasma pile-up in the outer heliosheath.  If the
partially ionized gas close to the Sun is in pressure equilibrium with
\Bis, then $P_{ \rm B} \sim P_{ \rm th} \sim 0.2$ eV and \Bis$\sim
2.8$ \muG.  If the gas and
interstellar magnetic field are in pressure equilibrium the
heliosphere has a Mach$\sim 1$ bowshock.

\section{Local Magnetic Field Direction from Interstellar Polarization Data}\label{sec:polarization}

The direction of the interstellar magnetic field at the Sun is
measured from starlight polarization.
Weak but systematic polarization of light from stars within 40 pc
in a region in the galactic center hemisphere was discovered by
\citet{Tinbergen:1982}.  Tinbergen attributed this weak polarization,
where detections are at the $3 \sigma  -5 \sigma$ level,
to a patch of ISM close to the Sun.
Tinbergen's data were acquired during solar minimum conditions,
in the south during 1974 and in the north 
during 1973 (J. Tinbergen, private communication).
\footnote{During the mid-1970's, the solar magnetic polarity was north
pole positive (A$>$0, field lines emerging at the north pole).}  
Paper I showed that this polarization reaches a maximum
at a location offset by $ \sim +35^\circ$
from the heliosphere nose.  Below it is shown that this
maximum also occurs in the ecliptic plane, and that the weak polarizations
extend to negative ecliptic latitudes
in the eastern hemisphere of the heliosphere (\S \ref{sec:data}).
The weak polarization observed by Tinbergen appears to arise
from interstellar dust interacting with the heliosphere (\S \ref{sec:heliosphere}).
Data are sparse and additional observations are needed.

\subsection{Polarization Data for Nearby Stars}\label{sec:data}

The catalog of polarizations of nearby stars used here consists of
data from \citet{Tinbergen:1982} and \citet[][acquired in
1974]{Piirola:1977}.  This combined Tinbergen-Piirola catalog (TPC)
has 202 stars, 86\% of which are within 40 pc (for Hipparcos distances).
The two data sets have 45 stars in common; the noisiest data from
the original studies are omitted from this discussion.  The errors on
the mean Stokes parameters are discussed by Tinbergen, and are $6 - 17
\times 10^{-5}$; combining different measurements yielded $1
\sigma \sim 6 \times 10^{-5}$ \citep[Table 1 of][]{Tinbergen:1982}.
The unit \P5$=10^{-5}$ degree of polarization will be used.  The
linear Stokes parameters U and Q give polarization \Pol =
(Q*Q+U*U)$^{1/2}$.  

\begin{figure}[h!]
\plotone{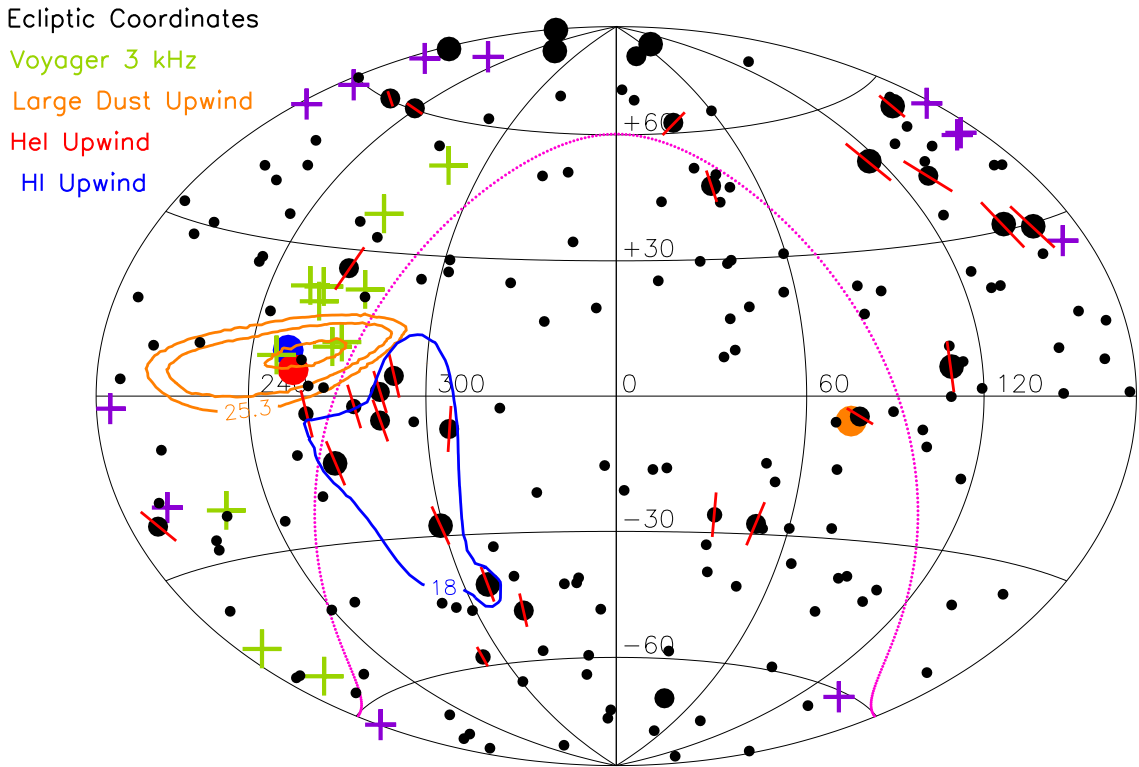}
\caption{The polarizations of the TPC stars within 45 pc are plotted
in ecliptic coordinates using an Aitoff projection.  The blue contour
outlines the locations of stars near $\lambda \sim 280^\circ$ with 
polarization $>0.018$\%, and the red
lines show the direction of the polarization plane of vibration.  The
heliosphere nose direction is defined by
the upwind directions of interstellar \HeI\ (red dot), and the \HI\ and downwind \HeI\ directions are
also shown (blue, and orange dots).  Orange
contours show the 1-$\sigma$, 2-$\sigma$, and 3-$\sigma$ uncertainties
on the upwind direction of large interstellar dust grains (typically
radii $ \gtrsim 0.2$ \micron) flowing into the heliosphere \citep{Frischetal:1999}. The
primary locations of the 3 kHz emission events detected by the Voyager
satellites are plotted as light green crosses, and the purple crosses
show the ambiguous alternate locations (Appendix \ref{app:khz}).
The pink dotted line shows the galactic plane.  The region of enhanced
polarization extends to negative ecliptic latitudes.
Small, medium and large dots indicate stars with polarizations of 15--19,
19--23, and over 23 $\times 10^{-5}$ degree of polarization,
respectively.  
\label{fig:aitoff} }
\end{figure}

The polarizations of the TPC are plotted in Fig. \ref{fig:aitoff} in
an Aitoff projection.  A region within 30\deeg\ of the ecliptic meridian 
$\lambda \sim 280^\circ$ shows the strongest polarizations down to latitudes of
$\beta \sim - 45^\circ$.  The polarization for stars close to 
the ecliptic plane reaches a maximum at a
location offset by $\Delta \lambda \sim +35^\circ$ from the heliosphere nose.

The average polarization as a function of $\lambda$ is plotted in Fig. \ref{fig:correl},
with averages smoothed over $\pm
20^\circ$ around the central ecliptic longitude \lambdazero\ because
of the small number of stars.  Averages are shown for stars within 20\deeg,
and within 50\deeg, of the ecliptic plane.  The polarization peaks at $\lambda \sim 295^\circ$
for stars with $| \beta | < 20^\circ$.
The polarization peak is offset by $\sim +30^\circ$ and $\sim +40^\circ$ from the
upwind directions of interstellar dust and \HeI, respectively.
The uncertainties on the polarization data have been carefully
carefully discussed by Tinbergen (1982) and Piirola (1977).
Paper I used only the Tinbergen (1982) data; the addition here of the Piirola (1977) data
improves the statistics for many regions of the sky, but not in the
heliosphere nose direction.  
Additional information on the smoothing process
is given in Appendix \ref{app:pa}.
These polarizations are weak, $\sim 0.02$\%, and have been detected at the $3 \sigma -5
\sigma$ levels.  
The polarization maximum regions coincides with
upstream direction of the ENA flux observed from the outer heliosphere
(\S \ref{sec:heliosphere}).
Fig. \ref{fig:aitoff} also shows the upstream direction of 
dust grains flowing into the heliosphere 
\citep[from Ulysses and Galileo data,][]{Frischetal:1999}
and of interstellar \HeI\ flowing into the heliosphere
\citep{Witte:2004,Moebiusetal:2004}.

The weak polarization becomes unobservable when the angle between the
sightline and the direction of the magnetic field, \Bhat, deviates
significantly from 90$^\circ$.  This condition can be estimated
because the observed polarization can not be
identified when $P < 2.5 \sigma$, so the polarization is only detected
in stars within $\sim 40^\circ$ of the perpendicular direction to
\Bis.  


\subsection{Is Polarization Origin at Heliosphere or in Nearby Upwind ISM?} \label{sec:origin}

\subsubsection{Polarization Origin at Heliosphere} \label{sec:heliosphere}

Several characteristics support the origin of this weak polarization
in the outer heliosheath region (defined as the region between the
heliopause and bow shock).  The first is that the upwind polarizations
are strongest in the ecliptic plane (\S
\ref{sec:data}).  The five stars that dominate the upstream
polarization maximum in the ecliptic plane have the same polarization
position angles (\S
\ref{sec:blocal}, Table \ref{tab:pa}, Appendix \ref{app:pa}).  
The polarization maximum region is elongated along the ecliptic plane
towards positive longitudes.
This may indicate either a blunt heliosphere or the sideways
deflection of the grains following \Bis\ around the heliosphere.

The second argument supporting an outer heliosheath origin of the
polarization is that the small grains capable of polarizing starlight
are tied by small gyroradii to the interstellar magnetic field
deflected around the heliosphere.  The magnetic field upstream of the
heliopause filters out grains with large charge-to-mass ratios, Q/M,
and gyroradii that are smaller than the characteristic lengths between
the heliosphere bow shock and heliopause
\citep[e.g.,][]{Frischetal:1999,CzechowskiMann:2003,LindeGombosi:2000}.
Dust grains in the LIC are silicates (SF07).  The radii of the
magnetically excluded grains are $\lesssim 0.1 - 0.25 $
\muG, and are comparable to the sizes of interstellar polarizing grains \citep[e.g.][]{Mathis:1986}.

The coincidence of the polarization maximum location with the ENA flux
from the outer heliosphere \citep{Collieretal:2004,Wurzetal:2004} may
indicate that maximum charge densities in the outer heliosheath occur
where the strongest compression of the interstellar magnetic field 
occurs.  The photoejection of electrons gives
positively charged dust grains \citep{Weingartner:2004}, and
ENAs form where interstellar neutrals charge exchange with protons, indicating
ENAs and small dust grains in the outer heliosheath may have
similar distributions.
The properties of the outer heliosheath regions are not yet understood,
and IBEX data on ENAs will elucidate the processes of the outer heliosphere
\citep{McComasetal:2005sw11}.

The polarization increases by $\sim 46$\% between the $3
\sigma$ detection towards 36 Oph AB at 6 pc, and HD 161892 at 39 pc
and 5\deeg\ away from 36 Oph AB.  This suggests the observed
polarization is partly formed beyond the heliosphere in
the cloud, but that the polarization level is boosted to detectable
levels in the ecliptic plane where the interstellar magnetic field
is compressed between the bow shock and heliopause and 
grain charging rates differ from those in the ISM.

Interstellar dust grains traversing the outer heliosheath
experience a gradient in magnetic field direction and strength, plasma
density, and grain charging rates, so that the grain gyroradii vary
with position. \citet{CzechowskiMann:2003} have modeled the passage of
interstellar dust grains across the transition region between the ISM and the heliopause
and found that in the presence of a strong shock small grains develop
velocity differences compared to the plasma velocity, and are
deflected to the heliosphere flanks.  Such a process may explain
the $\sim 35^\circ$ offset in the polarization maximum from the upstream
direction.  For some magnetic field directions and grain masses, grains are reflected back away from the
heliopause and form density waves.  The following section shows that
collisional disruption of grain alignment is slow.  This indicates that polarization
position angles will not vary if the \Bis\ component in the plane of the
sky is relatively constant as the field is initially displaced
by the heliosphere.

The polarizations appear to be strong for the small 
 ($\sim 150$ AU) distance between the heliopause and bow shock.
Comparisons between 36 Oph AB and HD 161892 suggests
that only half of the polarization must originate in the outer heliosheath
to boost the polarization strength to detectable levels (\S \ref{sec:data}).
There are no comparable data sets of high-sensitivity very weak optical
polarizations with which to compare the Tinbergen data.  

\subsubsection{Polarization Origin in Upwind ISM}\label{sec:upwind}

The Sun is located in an interstellar cloud that is part of a 
decelerating cluster of local interstellar cloudlets (CLIC) flowing away
from the direction of the Scorpius-Centaurus Association and past the
Sun \citep{Frisch:1995,FGW:2002}.  Figure \ref{fig:localfluff1} shows 
the distribution of neutral ISM within 15 pc for sightlines with 
\NHI$>1.0 \times 10^{18}$ \cmtwo, based on data in \citet{Woodetal:2005} 
and \citet{RLII}, and for some stars using the assumption  \DI/\HI$=1.5 \times 10^{-5}$.
Interstellar \HI\ within 15 pc is seen to coincide with the
eastern hemisphere of the heliosphere.  The initial interpretation of the
Tinbergen data was that the polarizations were formed in this nearby
patch of ISM in the galactic center hemisphere
\citep[T82;][]{FrischYork:1983,Bruhweiler:1984}. Comparisons between 
the polarizations of 36 Oph AB and HD 161892 suggest that the starlight
polarization observed in the ecliptic plane maximum also includes a contribution
from ISM upstream of the heliosphere nose (\S \ref{sec:data}). 
This ISM within 15 pc appears to be the closest part of a magnetic shell
from Loop I that has expanded to the solar location (see \S \ref{sec:loop}).

The polarization maximum in the ecliptic plane is seen to occur 
in a sightline through the boundary of the ISM within 15 pc.  
The possibility that the CLIC is the exclusive origin of the upwind
polarizations concentrated in the ecliptic plane can not be ruled out.
Most of the polarization is formed near the ecliptic meridian $\lambda \sim  280^\circ$,
which cuts through the low column density boundary of the ISM shown in Fig. \ref{fig:localfluff1}.
Ionization levels and magnetic field strengths may be higher 
towards the boundary regions than in the cloud interior, yielding stronger
polarizations.
Stars in the CLIC that are located between \el$\sim 330^\circ$ and \el$\sim 60^\circ$ 
are unpolarized.
For stars towards the polarization maximum near the ecliptic plane,
polarization position angles from interstellar dust grains in the outer
heliosheath and from grains in the upstream ISM should be the same,
since polarization position angles do not change significantly
over the nearest $\sim 200$ pc in this direction (\S \ref{sec:bheiles}),

Grain alignment persists over long time scales in low density gas such
as is upwind of the Sun, when compared to the denser clouds.  Low
densities significantly reduce the collisional disalignment of the
dust grains when compared to denser clouds.  The minimum and maximum
mean densities for the \HI\ within 15 pc are $<$\nHI$>$=0.01 \cc and
0.12 \cc, respectively, and the average density for stars within 15 pc
is 0.05 atoms \cc.  The LIC is less dense by a factor of $\sim$65,
warmer by a factor of $\sim$100, with proton densities a factor of
$\sim$10 larger than typical cold ISM.  Grain alignment is
disrupted when grains accumulate their own mass in thermal collisions
with the gas, which occurs over time scales of $\tau_\mathrm{mass} \sim
10^6 a \rho_\mathrm{gr} / n
\sqrt{T_\mathrm{gas} }$ Myrs = 0.7--1.2 Myrs in the LIC.  In the LIC
the collision rate is down by a factor of $\sim n / \sqrt{T} \sim 600
$ compared to denser clouds, so that even though \Bis\ is weaker in
the LIC compared to dense clouds, it is fundamentally easier to
magnetically align ISDGs in the LIC than in denser clouds.  Although
precise spatial densities are not known for the upwind ISM, the low
average densities indicate similar arguments will apply.


\section{Magnetic Field Direction at Polarization Maximum in Ecliptic Plane}\label{sec:bfield}

The direction of the local interstellar magnetic field is established
by polarization position angles of starlight.  The magnetic field
direction towards the polarization maximum in the ecliptic plane is the same
for local stars (\S \ref{sec:blocal}) and distant stars in the same
region (\S \ref{sec:bheiles}).  The polarization position angle in 
celestial coordinates of the plane of vibration of the electric vector 
is given by \thetac=0.5 arctan(U/Q) (also see Appendix
\ref{app:pa}).  The interstellar magnetic field is parallel to the
polarization, as is shown by synchrotron emission that is polarized
perpendicular to the field direction \citep[e.g.][]{Heiles:1976araa}.

\subsection{Magnetic Field Direction near the Sun}\label{sec:blocal}

The stars in the polarization maximum near the heliosphere nose have
a mean position angle in ecliptic coordinates of \PAecl $=-25^\circ \pm 4^\circ$ (Table \ref{tab:pa}).  
The polarization position angles
do not vary significantly with ecliptic longitude in the
polarization maximum region (see Appendix \ref{app:pa}).
The interstellar magnetic field direction is parallel to the polarization vector
\citep{Spitzer:1978}.
The polarization position angles indicate that the
magnetic field direction close to the Sun 
is inclined by $\sim 55^\circ$ with respect to the
galactic plane and $\sim 65^\circ$ with respect to the ecliptic plane,
regardless of whether the polarization originates in the outer
heliosheath or in the very local ISM.

The 5\deeg\ angular offset between the interstellar \HeI\ and \HI\
flowing through the heliosphere, discovered by
\citet{WellerMeier:1974} and now accurately determined using Ulysses
\HeI\ data and SOHO \HI\ \Lya\ florescence data, is also a diagnostic
of the interstellar magnetic field direction at the Sun
\citep{Witte:2004,Quemeraisetal:2000,Lallementetal:2005}.  The \HI\ and
\HeI\ upstream directions (Table \ref{tab:direct}) can be used to
define a position angle for comparison with the polarization data.
This position angle is the \HI\ upwind direction with respect to a great
circle meridian passing through the \HeI\ upwind direction, and is
PA$_\mathrm{E,HHe} = -33^\circ \pm 14^\circ$ in ecliptic coordinates
(Table \ref{tab:pa}).  
The magnetic field directions defined by starlight polarization in the
heliosphere nose, and by the 5\deeg\ offset between \HI\ and
\HeI\ flowing into the heliosphere, therefore give the same result to within the uncertainties.
However a component of the magnetic field that is parallel to the sightline
would not be detected by these methods.

\subsection{Magnetic Field Direction from Distant Stars} \label{sec:bheiles}

The upwind direction of the LIC corresponds to an astronomically
interesting direction, dominated by the North Polar Spur
\citep[][]{Frisch:1981}, which is the radio-intense section of the
Loop I superbubble shell that is traced by both polarized synchrotron
emission and magnetically aligned dust grains
\citep[e.g.][]{Heiles:1976araa}.  As a test of the relation between
the nearby and distant magnetic field in the direction of the
heliosphere nose, the polarization position angles for stars in the
TPC are compared with those of distant stars in the Heiles
Polarization Catalog \citep[HPC,][available on Vizier as catalog
II/226]{Heiles:2000pol}.  The scaled polarizations for stars 40--100
pc distant and within 20\deeg\ of the ecliptic plane are plotted
versus \el\ in Fig. \ref{fig:correl}, where the polarizations are
smoothed over ecliptic intervals of $\pm 5^\circ$ from the central
longitude.  Locations where the ecliptic crosses Loop I are seen as
two polarization peaks at \el$\sim 264^\circ$ and $\lambda \sim
220^\circ$, and the strongest peak near \glong$\sim 4^\circ$ and
\glat$\sim 4^\circ$ is within 10\deeg\ of the ecliptic nose direction.

The polarizations of stars in the HPC within 250 pc are displayed in
Fig.  \ref{fig:loop}, using only objects with HD numbers and
parallaxes in the Hipparcos catalog \citep{Perrymanetal:1997}.  No
screening is made for discrepancies between the celestial and galactic
positional angles listed in the HPC.  The familiar pattern of magnetic
filaments traced by polarization vectors is evident.

The comparison between the position angles of near and far stars in
the heliosphere nose region of maximum polarization has two distinct
properties.  First, for stars in the heliosphere nose region the
average position angle of the distant stars in the HPC is
\PAecl$=-20^\circ \pm 20^\circ$, in agreement with the TPC data (Table \ref{tab:pa},
Fig. \ref{fig:loop}).  Slightly different definitions of the
heliosphere nose region are used for the TPC versus the HPC stars
because of the small number of TPC stars (see Table \ref{tab:pa} for
region specifications).  These position angles for the nearby TPC
stars agree to within uncertainties with those of the distant HPC
stars in the region of maximum nearby polarization.  For example, the
nearby star HD 155885 (6 pc) and distant star HD 157236 (189 pc) in
the same direction towards the heliosphere nose both have the same
polarization position angle (\PAgal=39\deeg\ in galactic coordinates), although the
polarization of HD 157236 is stronger than for HD
155885.  Second, for a more extended spatial interval around
the heliosphere nose the position angles of the distant stars vary
systematically with \emph{ecliptic} latitude, due to the chance
alignment of a Loop I magnetic filament (this is shown in Appendix
\ref{app:pa}).  

Close to the ecliptic plane
in the polarization maximum region the distant and nearby stars
show the same mean polarization position angles.  The similarity in the polarization position angles of
nearby stars within 40 pc, and distant stars (140--240 pc) that trace
the Loop I magnetic field, indicates that 
the interstellar magnetic field is
well-ordered over several hundred parsecs in this direction,
and interstellar densities are low, minimizing 
the collisional disruption of the aligned grains.
It is suprising that the polarization data indicate the interstellar
magnetic is well-ordered over a few hundred parsecs, especially since
the distant region includes the Riegel-Crutcher cold cloud where
magnetic filaments are seen
\citep{RiegelCrutcher:1972,McClurGriffiths:2006}.  However 
the Local Bubble surroundings of the Sun 
provide a minimal barrier to the expansion of Loop I \citep{Frisch:1981}.
Superbubbles formed by stellar evolution in
the Sco-Cen Association during the past $\sim 5$ Myrs evidently
dominate morphology of the local interstellar magnetic field.
\section{Evidence for Unrecognized Foreground Contamination of the Cosmic Microwave Background }\label{sec:cmb}

Paper I suggested that the interstellar dust grains trapped in the
outer heliosheath and deflected around the heliosphere may contribute
a weak unrecognized foreground to the cosmic microwave background.
The large scale CMB power spectrum contains poorly understood
anomalies related to ecliptic geometry, indicating that the low-$\ell$
CMB may have unrecognized very local foreground contamination related
to the ecliptic geometry
\citep[e.g.][]{Eriksenetal:2004,SchwarzStarkmanetal:2004}.  These
anomalies appear in the Year 1, and remain in the Years 1-3 WMAP data
after instrumental corrections were improved
\citep[][]{HinshawNoltaetal:2007,Copietal:2007}.
The interstellar magnetic field near the Sun (\S \ref{sec:bfield}) and
small interstellar dust grains excluded from the heliosphere (\S
\ref{sec:data}) both provide possible foreground contaminants of the
CMB radiation.

In this section the geometrical properties of the CMB low-$\ell$
moments are compared to the geometrical properties of heliospheric
phenomena.  Several coincidences are found.  The geometrical
properties of what has been identified as the Doppler contribution to
the CMB dipole moment \citep[e.g.][]{Matheretal:1994} are related to
both the heliosphere geometry and the distribution of ISM within 35
pc.  CMB low-$\ell$ multipole moments, including the non-cosmological
Doppler shift to the CMB dipole moment, and the quadrupole and octopole
multipoles, trace important heliospheric quantities.  
The interstellar magnetic field at the Sun may influence these apparently independent
phenomena, and this field direction appears to be determined
by the S1 subshell of Loop I (\S \ref{sec:loop}).

This qualitive discussion focuses on the spatial coincidences
between the CMB and heliosphere properties.  The source of possible
heliospheric contamination of the CMB microwave radiation is unknown.
The outer heliosphere region is only recently explored for the first
time by Voyager 1 and Voyager 2, which have crossed the solar wind
termination shock at 94 AU and 85 AU in December 2004 and August 2007,
respectively.  Since the heliosphere itself varies with the magnetic
solar cycle of the Sun, any possible heliospheric foreground to the
CMB radiation may vary with solar cycle phase.

\subsection{CMB Features Related to Heliosphere and Local ISM }\label{sec:cmbhelio}

Figures \ref{fig:copi} and Fig. \ref{fig:ilc3} summarize the geometrical
coincidences between the CMB low-$\ell$ multipole moments and
heliospheric properties.  

Structure in the Cosmic Microwave Background (CMB) signal over large
angular scales was detected in the COBE maps
\citep[][]{Smootetal:1992,KogutSmootetal:1992,Matheretal:1994}, and
persists in the high signal-to-noise Wilkinson Microwave Analyzer
Polarimeter (WMAP) data
\citep[e.g.][]{BennettHilletal:2003,HinshawNoltaetal:2007}.  The
anisotropies of the CMB at large angular scales are dominated by
characteristic anisotropies which include a dipole moment of amplitude
$\Delta T/T \sim 10^{-3}$ due to the energy shift induced by the
Doppler motion through the CMB rest frame, a weaker quadrupole moment
with amplitude $\Delta T/T \sim 10^{-5.2}$, and weak higher order
asymmetries
\citep{Smootetal:1977,Smootetal:1992,KogutLineweaveretal:1993}.

The plane that separates the hot and cold hemispheres of the CMB
dipole moment traverses the heliosphere nose $\sim 6^\circ$ from the
upstream direction defined by interstellar \HeI\ flowing into the
heliosphere (Table \ref{tab:direct}).  The poles of the CMB Doppler
motion are therefore nearly equidistant from the inflow direction of
ISM into the heliosphere, and the plane separating the hot and cold
hemispheres of the CMB dipole moment separates the eastern and western
parts of the heliosphere (Fig. \ref{fig:copi}).
\footnote{This fact does not appear to have been discussed
previously in the literature.
\citet{Smootetal:1977} report that Peebles referred to the
CMB kinematic Doppler dipole as the ``new aether drift''.}  The
position angle defined by this plane, at the point of closest approach
to the heliosphere nose, is within 15\deeg\ of the position angles of
the polarizations in the ecliptic plane maximum (Table
\ref{tab:pa}).
\footnote{By coincidence the heliocentric upstream direction of ISM
flowing through the heliosphere, which represents the vector sum of
the solar and LIC motions through the LSR, is within $\sim 5^\circ$ of
the ecliptic plane and $\sim 15^\circ$ of the galactic center.}
\nocite{Copietal:2007} 
The plane separating the hot and cold hemispheres of the CMB dipole
moment therefore appears to be within $ \sim 15^\circ$ of the
interstellar magnetic field direction at the Sun found from optical
polarizations.  For comparison, the WMAP ILC3 map data
\citep{HinshawNoltaetal:2007} are plotted in Fig. \ref{fig:ilc3} in
ecliptic coordinates, with the same features as shown as in
Fig. \ref{fig:copi}.

Anomalies also appear in the low-$\ell$ quadrupole and octopole
moments.  Using ``multipole vectors'' to decompose the Internal Linear
Combination map, Years 1-3 (ILC3) data, Copi et al. (2007) found that
there is an alignment between the quadrupole and octopole planes, and
that the ecliptic traces a zero between the extremes in the quadrupole
and octopole moments.  The multipole vectors carry the angular
distribution information of the power for each $\ell$-value.  Copi et
al. showed that the $\stackrel{\rightarrow}{w}$ area vectors (normals
to planes defined by multipole pairs) coincide with the ecliptic plane
for southern galactic latitudes.
In Fig. \ref{fig:copi} the positions of the $\ell=$2 and 3 multipole
vectors $\hat{v}$, and $\stackrel{\rightarrow}{w}$ area vectors, are
plotted in ecliptic coordinates \citep[locations from ][Table 1,
ILC123 data]{Copietal:2007}.  The $v^{22}$ component is directed
towards the heliosphere nose.  The four area vectors $w^{323}$,
$w^{212}$, $w^{312}$, and $w^{331}$ are all directed towards or close
to the strip on the sky containing the 3 kHz emissions (Appendix
\ref{app:khz}) and the hot pole of the dipole.  This strip is nearly
perpendicular to the ecliptic plane, in the sidewind direction
corresponding to ecliptic longitude $\lambda \sim 180^\circ$
(Fig. \ref{fig:aitoff}).  The 3 kHz emissions are formed where outward
propagating global merged interaction regions interact with the
magnetically-shaped heliopause \citep{MitchellCairnsetal:2004}.  The
low-$\ell$ moments moments of the CMB therefore appear to respond to
the ecliptic coordinate system partly because of the shaping of the
heliosphere by the interstellar magnetic field.

Separate analysis of the WMAP data as a function of solar cycle phase
may yield clues about the origin of the new foreground.  Copi et
al. (2007) pointed out that there is a discrepancy between the Year 1
and Year 1--3 quadrupole moments.  WMAP was launched in June 2001, and
the Year 1 data represent data acquired between Sept. 2001 and
Sept. 2002 during solar maximum conditions, versus between Sept. 2001
and Sept. 2004 for the Year 1--3 period that includes the solar
minimum conditions of 2004.  The shift of the $v^{22}$ multipole from
\el=268.9\deeg, \eb=15.6\deeg\ during Year 1 (17.6\deeg\ away from the
upwind direction) to \el=252.6\deeg, \eb=8.2\deeg\ during Years 1--3
(3.7\deeg\ from the upwind direction) indicates that the process
contributing CMB foreground may be solar cycle dependent.  An
analogous shift has not been reported for the upstream direction of
inflowing \HeI, which has been measured up to 2003 by Ulysses
\citep{Witte:2004,Moebiusetal:2004}.


\subsection{The CMB Dipole Moment and Local Interstellar Matter}\label{sec:cmbclic}

The cold hemisphere of the CMB dipole moment coincides 
with the location of neutral ISM within 15 pc with \NHI$>10^{-18}$ \cmtwo\ (Fig. \ref{fig:localfluff1}).  
When the highest column densities of interstellar \HI\ within 35 pc
are plotted, the hot pole of the CMB dipole
moment is seen to coincide with distant parts of the S1 magnetic
shell (Fig. \ref{fig:localfluff2}, \S \ref{sec:loop}). 
Together with the polarization data,
this indicates that the S1 magnetic bubble has expanded to the
solar location.  CMB foreground templates calculated for synchrotron emissivity 
do not include contributions from a local magnetic field
or this magnetic bubble.

\section{S1 Subshell of Loop I Superbubble as Explanation of Local Magnetic Field }\label{sec:loop}

The relation between the ISM at the Sun
and the Loop I superbubble formed by stellar evolution in the Scorpius-Centaurus
Association was discovered from the kinematics and abundances seen
in ISM upwind of the Sun \citep{Frisch:1981}, and has now been investigated 
in a number of studies
\citep[e.g.][]{Crutcher:1982,Bochkarev:1984,Crawford:1991,deGeus:1992,Frisch:1995,Breitschwerdtetal:2000,MaizApel:2001}.
Recently the Loop I magnetic superbubble has been modeled as the
superposition of two magnetic bubbles with centers offset from each
other by $\sim 37^\circ$ \citep{Wolleben:2007}.  The nearest of these
structures, S1, is centered at \glong$=346^\circ \pm 5^\circ$,
\glat$=3^\circ \pm 5^\circ$, distance $68 \pm 10$ pc, with radii for
the inner and outer rims of $72 \pm 10$ pc and $91 \pm 10$ pc,
respectively.  The inner and outer tangential rims 
of S1 are approximated as circles and plotted in
ecliptic coordinates in Figs. \ref{fig:localfluff1} and \ref{fig:localfluff2}, and in galactic coordinates in Fig. \ref{fig:loop}.  
The nearest CLIC gas, $d < 15$ pc, 
coincides roughly with the eastern part of the S1 shell (Fig. \ref{fig:localfluff1}). 
Interstellar \HI\ with \NHI$> 2 \times 10^{-18}$ traces the tangential regions through
the western part of the shell (Fig. \ref{fig:localfluff2}).  The
upwind direction of the CLIC in the local standard of rest
is centered near the center of the S1
shell.  
The best fit bulk flow velocity of interstellar cloudlets within 30 pc is 
$-17 $ \kms\ from \glong,\glat = 0\deeg, --5\deeg\ \citep{FGW:2002},
using the Hipparcos solar apex motion \citep{DehnenBinney:1998}.
 
The CLIC \HI\ densities decrease towards the shell center
as if the CLIC is part of the expanding shell.  The \HI\ minimum is
centered near \glong$\sim 330^\circ$ and \glat$\sim +8^\circ$, where
the radiation field from hot stars from the Upper Centaurus Lupus
subgroup of the Centaurus association appears to ionized upwind CLIC
gas in the solar vicinity.  An additional source of
ionizing radiation is the white dwarf star WD1634-573, located
at \glong=330\deeg, \glat=--7\deeg, and  where 60\% of the gas is ionized
\citep{Woodetal:2002}.

The locations of the two poles of the CMB dipole moment both fall just outside of the S1
shell, and coincide with the \HI\ gas surrounding the S1 magnetic shell (Fig. \ref{fig:localfluff2}).

Partially ionized ISM is tied to the interstellar magnetic field so
that gas and dust are trapped in the expanding S1 magnetic shell.  If
the LIC gas and magnetic field at the Sun are in pressure equilibrium,
then $|$\Bis$| \sim 2.8$ \muG\ (Model 26, SF07).  For the warm LIC, $T
> 5000$ K, the proton gyroradius is $< 0.4$ pc and the ISM will couple
to the magnetic field in the moving shell unless cloud densities are
\nHI$> 2$ \cc\ \citep[which is possible but not
likely,][]{Frisch:2003apex}.  However there is a well-known
discontinuity in the velocity of the LIC versus the upstream gas
towards the nearest star $\alpha$ Cen
\citep[e.g.][]{LinskyWood:1996,Landsmanetal:1984}, indicating possible
substructure in the S1 shell induced by higher field strengths.  The
position angle data indicate that the magnetic field direction does
not vary significantly over several hundred parsecs in the upstream
direction in the region of the ecliptic-plane-polarization-maximum, so there must be empty regions of space with minimal dust
between the near and far fields, such as expected for the interior of
a magnetic bubble.

Rotation measures of pulsars and extragalactic radio sources show that
the uniform component of the interstellar magnetic field in the
interarm region around Sun is directed towards \glong$\sim 82^\circ$.
The strength of the large-scale galactic regular magnetic field at the
solar radius is $B_\mathrm{IS} \sim 2.1 \pm 0.3 $ \muG, found from
rotation and dispersion measures towards an assembly of pulsars within
several kiloparsecs of the Sun \citep{Han:2006}.  Faraday rotation data
show that the large-scale magnetic field reverses polarity around the
galactic \glong=0\deeg\ meridian at the galactic center and also
around the galactic plane.  In the part of the CLIC within 15 pc,
\glat$<0^\circ$ and \el$>270^\circ$, the global field polarity is directed
away from the Sun.  

The polarization data establish the direction of the magnetic field
at the Sun (Table \ref{tab:pa}).  Radio synchrotron polarization data
indicate that the optical polarization vector is parallel to the
magnetic field direction in the Loop I filament, and there is no
reason to assume otherwise for the local ISM, especially given the
coincidence in position angles for near and far polarization.
However, there may be a component of \Bis\ that is perpendicular to
the plane of the sky that is not sampled by optical polarization.  For
a high-latitude direction through the distant Loop I, the \Bis\ component parallel
to the sightline is small.  At the location \glong$= 34^\circ$,
\glat$=42^\circ$, where a $\sim 70 \pm 30$ pc tangential sightline
through Loop I is sampled, the volume-averaged field strength is \Bis$
\sim 4$ \muG\ \citep{Heilesetal:1980}.  For this sightline, the
Faraday rotation of extragalactic radio sources indicate the \Bis\
component parallel to the sightline is small, with an average value of
$ B_\mathrm{||} = 0.9 \pm 0.3 $ \muG\ \citep{Fricketal:2001}.


\section{Conclusions}  \label{sec:conclusions}
\begin{itemize}

\item 

The polarizations of nearby stars, $< 45 $ pc, form a maximum
in the ecliptic plane that peaks at an ecliptic longitude shifted by
$\sim 35^\circ$ from the upstream direction of interstellar gas and
dust flowing into the heliosphere.  Half of this polarization is
formed within $\sim 6$ pc of the Sun, and the polarization maximum
indicates a significant contribution from
small interstellar dust grains interacting with the outer
heliosheath regions.  

\item

The polarization position angles of nearby and distant stars in this
polarization peak agree to within the uncertainties (Table
\ref{tab:pa}).  The position angle defined by the offset between the
upstream directions of interstellar \HeI\ and \HI\ in the heliosphere
also agrees with the position angles found from the polarization-peak
stars, to within the uncertainties.  

\item

These polarization data give a direction for the
interstellar magnetic field at the Sun that is inclined by $\sim
65$\deeg\ with respect to the ecliptic plane, and $\sim 55$\deeg\ with
respect to the galactic plane.  An additional component into the plane
of the sky may also be present.  If the interstellar magnetic field
direction nearby is similar to the global field in the solar vicinity,
then the field is directed away from the Sun east of the heliosphere
nose in ecliptic coordinates.

\item

The plane that separates the cold and hot hemispheres of the CMB
dipole moment also divides the east and west hemispheres of the
heliosphere, and also encloses most sightlines within 15 pc that have
interstellar \NHI$>1.0 \times 10^{18}$ \cmtwo.

\item The CMB quadrupole area vectors of \citet{Copietal:2007} line up
with the positions of the locations of the 3 kHz events detected by
Voyagers 1 and 2.  A mix of primary and alternate locations for the 3
kHz emissions are used for this comparison (Appendix \ref{app:khz}).

\item

A plane can be formed from points that are equidistant between the 
two poles of the CMB dipole.
At the closest point to the heliosphere nose
direction defined by inflowing interstellar \HeI, the position angle of this plane is within $\sim 15^\circ$ 
of the position angles of the optical
polarizations that form the ecliptic-plane-polarization-maximum.

\item

\citet{Wolleben:2007} has modeled the Loop I magnetic field in terms
of two bubbles, one of which (S1) has expanded to the solar location.
The distribution of interstellar \HI\ within 35 pc shows evidence of a
shell morphology coinciding with the rims of S1.  The eastern part of
the shell, in ecliptic coordinates, is closest to the Sun.  The
upstream direction of the ISM flow past the Sun is towards the center
of the S1 shell.  The polarizations seen by Tinbergen are somewhat concentric
with the curvature of the S1 shell for southern ecliptic regions.

\item

The hot and cold CMB Doppler dipole moment poles are both directed
toward the boundaries of the S1 shell.

\item 

These results suggest that the magnetic field associated with
the S1 subshell of the Loop I supernova remnant  
creates the interstellar magnetic field at the heliosphere nose.
The S1 shell dominates the polarization of stars close to the Sun,
the distribution of ISM within 15 pc, the heliosphere
configuration that yielded the 3 kHz emissions, and the hot 
and cold poles of the CMB Doppler dipole moments.

\item

The CMB foreground that mimics the spatial
distribution of the 3 kHz emissions and the east-west heliosphere
symmetry may arise from the heliosphere itself,
or from another property that is also affected by the 
interstellar magnetic field in the S1 shell, such as the spectrum of synchrotron emission or dust grains trapped and heated in the shell.

\item

The polarization maximum in the ecliptic plane is dominated by five stars
with the same polarization position angles.  The polarization detections
by Tinbergen (1982) for these stars are $3 \sigma - 5 \sigma$, and
additional data on the polarization of nearby stars are desirable.
Tinbergen's data were collected during solar minimum in the mid-1970's,
and there may be a solar cycle dependence to the polarization strength.

\end{itemize}

\acknowledgements This research has been supported by the NASA the
grants NAG5-13107 and NNG05GD36G to the University of Chicago.  I
would like to thank Dragan Huterer for important comments on the CMB
multipoles, Carl Heiles for helpful suggestions, and Hiranya Peiris,
Cora Dvorkin, and Bruce Winstein for providing the ILC map data for
Years 1-3 in {\sc ascii} form.

\appendix

\section{Additional Information on the Polarization Data}\label{app:pa}

\begin{figure}[!ht]
\plotone{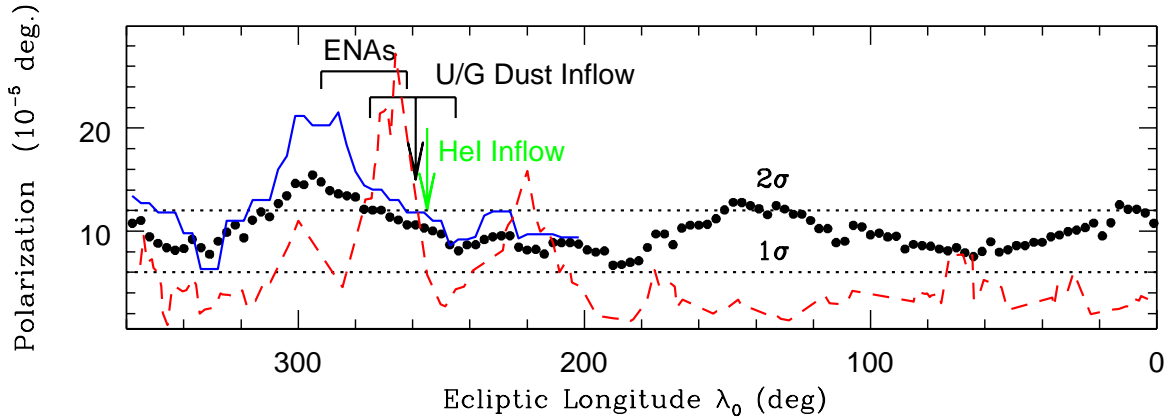}
\vspace*{-4.4in}
\caption{ Polarization properties as a function of ecliptic longitude:
The blue line shows the averaged polarizations \Pol\ for stars
with $| \beta| <$20\deeg, and the dots indicate averaged values for
stars with $| \beta| <$50\deeg\ in the TPC.  The $1 \sigma$ and $2
\sigma$ polarization uncertainties are shown (from Tinbergen 1982).
The polarization data for nearby stars in the TPC are averaged over
$\pm 20$\deeg\ in ecliptic longitude, $\lambda$.  The direction of
maximum \Pol\ is shifted by $\sim 30^\circ $ from the upwind direction
of interstellar dust grains flowing through the heliosphere, and by
$\sim 40^\circ $ from the \HeI\ upwind direction.  The black arrow
shows the upstream direction, and $1 \sigma$ uncertainties, for
interstellar dust flowing through the heliosphere based on Ulysses and
Galileo data (F99).  The green arrow shows the upstream direction from
observations of interstellar \HeI\ in the inner heliosphere
\citep[from][]{Witte:2004}.  The upwind direction of the ENA flux
originating in the outer heliosphere is indicated
\citep[from][]{Collieretal:2004,Wurzetal:2004}.  The red dashed line
shows the relative polarizations of 184 stars in the HPC
\citep{Heiles:2000pol}, with $| \beta | < 20^\circ$ and 40--100 pc
from the Sun.  The polarization strengths of the HPC data are reduced
by a factor of 20 for comparison with the TPC data, and are averaged
over an interval of $\pm 5^\circ$ around the central ecliptic
longitude.  The two strongest maximum in the HPC data are where the
sightline crosses the North Polar Spur (strongest peak,
\glong,\glat$\sim 4^\circ, 4^\circ$) and towards an \HI-21 cm filament
that is part of the Egger interaction ring \citep[\glong,\glat$\sim
336^\circ, 42^\circ$,][]{Egger:1995}. 
\label{fig:correl} }
\end{figure}

The details of the statistics of the groups of stars averaged together
to create Fig. \ref{fig:correl} are as follows: \emph{For the group of
129 stars that are within 50\deeg\ of the ecliptic plane and 40 pc of
the Sun:} A. Between 7 and 22 points contribute to the individual
stepped central average ecliptic latitudes ($\lambda_0$).  B.  For the
11 data points in the polarization maximum region in the interval
$\lambda \sim 280 - 310 $\deeg, between 11 and 15 stars contribute to
the averaged polarization displayed at each stepped $\lambda_0$ in
ecliptic longitude.  C.  In this interval, the averaged polarizations
are $11.4 - 15.5 \times 10^{-5}$ degree of polarization.  D.  The rms
variation of these averaged polarizations is $\lesssim 8.4 \times
10^{-5}$ degree of polarization.  \emph{For the group of 51 stars that
are within 20\deeg\ of the ecliptic plane and 40 pc of the Sun:} A.
Between 2 and 10 stars contribute to each data point, and 2--6 stars
contribute to each data point in the $\lambda \sim 280 - 310 $\deeg\
interval of the polarization maximum.  B.  In this interval, the
averaged polarizations are $13.0 - 21.5 \times 10^{-5}$ degree of
polarization.  D.  The rms variation of these averaged polarizations
is $\lesssim 8.5 \times 10^{-5}$ degree of polarization.

Fig. 5 of \citet{Tinbergen:1982} shows the statistical significance of the
individual data points contributing to this plot, displayed in
galactic coordinates (with distances based on pre-Hipparcos
distances).

The position angle of a vector is defined with respect to the
north-south meridian passing through the position of the object, so
that PA increases in the direction of increasing longitude.
In this paper, position angles are calculated using the IDL routine posang.pro.
Position angle transformation between coordinate systems were
calculated using unweighted averages of U and Q, using the IDL
function atan(U,Q), which keeps track of the quadrant.  If
$\theta_\mathrm{G}$ (or $\theta_\mathrm{E}$) is the PA in the galactic
(or ecliptic) coordinate system, and $\ell_\mathrm{N},~ b_\mathrm{N}$
are the galactic (or ecliptic) coordinates of the equatorial system
north pole, then from Appenzeller (1968), e.g.,
\begin{equation}
cot(\theta_\mathrm{C}-\theta_\mathrm{G}) = \frac{cos(b)*tan(b_\mathrm{N}) - 
	cos(\ell-\ell_\mathrm{N})*sin(b)}{sin(\ell -\ell _\mathrm{N})}
\end{equation}


The \citet{Tinbergen:1982} and \citet{Piirola:1977} results are
consistent.  Tinbergen (his Table 4) found that the mean differences
in the Q and U values between the Tinbergen and Piirola surveys, for
overlapping stars, are $ 0.3 -0.6$ and $\sim 2.1 -2.3$, respectively,
in units of \P5$= 10^{-5}$ degree of polarization, indicating the two
data sets are in good agreement.

The stars dominating the polarization maximum within 20\deeg\
of the ecliptic plane and near $\lambda \sim 295^\circ$ 
figure HD 155885, HD 161892.  HD 169916, HD 177716, and
HD 181577 (Table \ref{tab:star}).
The high ecliptic latitude stars with strong polarizations are
HD 137759, HD 142373, HD 150997, HD 153597, HD 163588, HD 170153, HD 185144, and HD 216228.
The statistically insignificant polarization maximum seen near
\lambdazero$\sim$140\deeg\ in Fig.  \ref{fig:correl} is dominated by
the three high galactic latitude stars HD 90839, HD 95689,
and HD 98230.
The stars with \P5$>3 \sigma$ at low ecliptic latitudes are
HD 205478, HD 196171, HD 209100, and HD 219571, where
\P5=$P/10^{-5}$ degree of polarization.  


For a larger spatial interval around the heliosphere nose, the
position angles of the distant stars are found to vary systematically
with \emph{ecliptic} latitude.  This variation is due to a
coincidental alignment of a Loop I magnetic filament with \eb, and is
a byproduct of the curvature of a Loop I magnetic filament that
extends between \glong$\sim 330^\circ$ and \glong$\sim  - 345^\circ$, at $b \sim
-12^\circ$.  The distance and ecliptic latitude dependencies of
the galactic position angle \PAgal\ are plotted for stars in the vicinity of the heliosphere nose
in Fig. \ref{fig:poly}.  The TPC stars are in the interval $ 255^\circ
< \lambda < 340^\circ $ and $ -35^\circ < \beta < 10^\circ $, and the
HPC stars are additionally restricted to the interval $ -18^\circ < b
< 16^\circ $ and either $ \ell > 355^\circ$ or $ \ell < 5^\circ$.  The
PA of HD 152424 is anomalous and the star is omitted from the plots.
Fig. \ref{fig:poly}, top, shows that there is no evidence for a systematic
distance dependence of \PAgal\ with the distance of the star.
In contrast, \PAgal\ rotates towards the east (larger values) at lower ecliptic
latitudes (bottom).  This coincidental rotation of \PAgal\ with ecliptic
latitude shows the curvature of the magnetic field traced by distant
stars that sample a filament of the Loop I supernova remnant.  The best-fit first
order polynomial, \PAgal$=a - b \beta$, for the 26 HPC and TPC stars 
in this region with $D < 400$ pc and
\eb$> -35^\circ$  yields $a = 8.67 \pm 4.54$ and $b = -2.19 \pm 0.24$
(blue line, Fig. \ref{fig:poly}, bottom).  For stars within 200 pc,
\PAgal\ values can be fit with a polynomial 
with $a = 28.57 \pm 5.35$ and $b = -0.56 \pm 0.74$ (orange line, 
Fig.  \ref{fig:poly} bottom).  Polarization measurements of additional 
nearby stars are required in the heliosphere nose direction to improve
the comparisons between the position angles of near and distant
stars. 


\section{Voyager 3 kHz Emissions} \label{app:khz}

\begin{figure}[h!]
\plotone{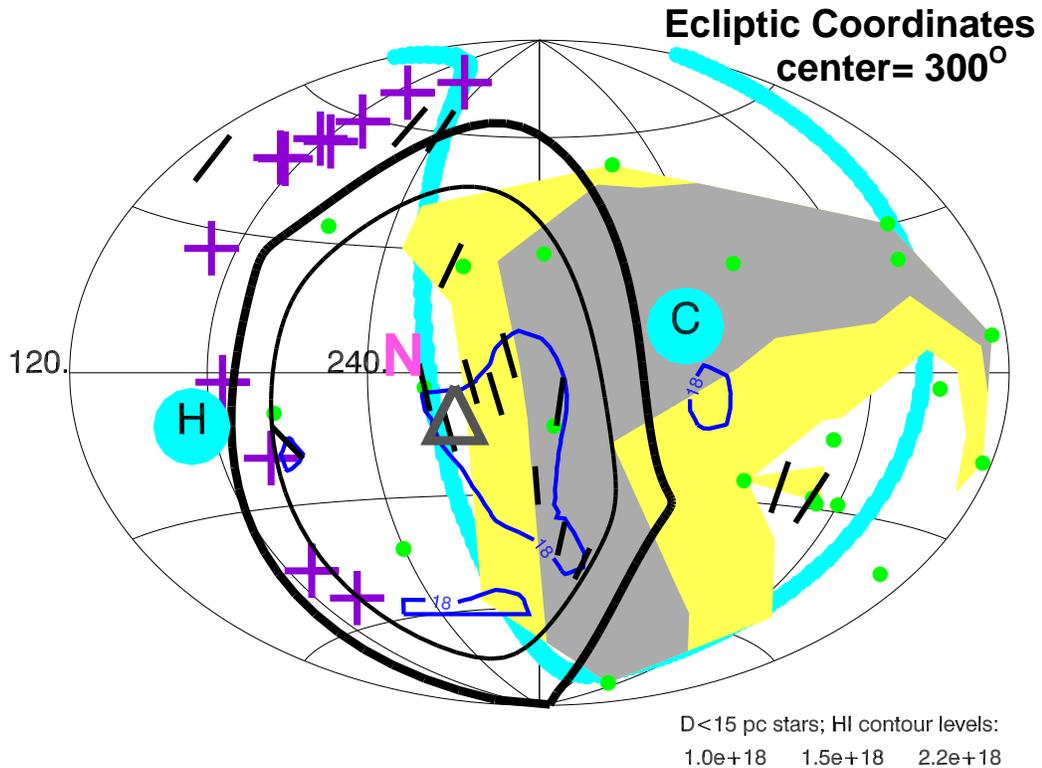}
\caption{The distribution of interstellar \HI\ within 15 pc is
plotted in the ecliptic coordinate system.
The plot is centered at $\lambda =
300^\circ$.  Stars within 15 pc are used to define the
distribution of local ISM with \NHI$> 1 \times 10^{18}$ \cmtwo.  The
polarizations of stars in the TPC with
\P5$> 18$ are plotted as black bars, and the
dark blue contour line shows the contour around these values.  The
levels of the filled yellow and gray \NHI\ contours are listed in the
plot, and green dots show the positions of the stars used
to define the contours.  The data are from \citet{Woodetal:2005} and
\citet{RLII}, with D/H=$1.5 \times 10^{-5}$ in some
cases.  Large cyan-colored dots with "H" and
"C" over-plotted show, respectively, the hot and cold poles of the
CMB Doppler dipole moment.  The cyan-colored line divides the hot and
cold hemispheres of the CMB Doppler dipole moment.  The purple crosses
show the locations of the 3 kHz emissions seen by the
Voyager satellites (see text and Appendix \ref{app:khz}).  The
large concentric "circles" show the inner and outer boundaries of the
"S1" magnetic loop, which is part of the Loop I magnetic bubble
that has expanded to the solar location \citep{Wolleben:2007,Frisch:1981}.  The gray triangle
shows the upwind direction of the CLIC in the LSR for an assumed solar
apex motion corresponding to the Hipparcos-based value
\citep{FGW:2002}.  The heliosphere nose direction, according to
interstellar \HeI\ flowing into the heliosphere, is marked by the
large pink "N".
\label{fig:localfluff1} }
\end{figure}

\begin{figure}
\plotone{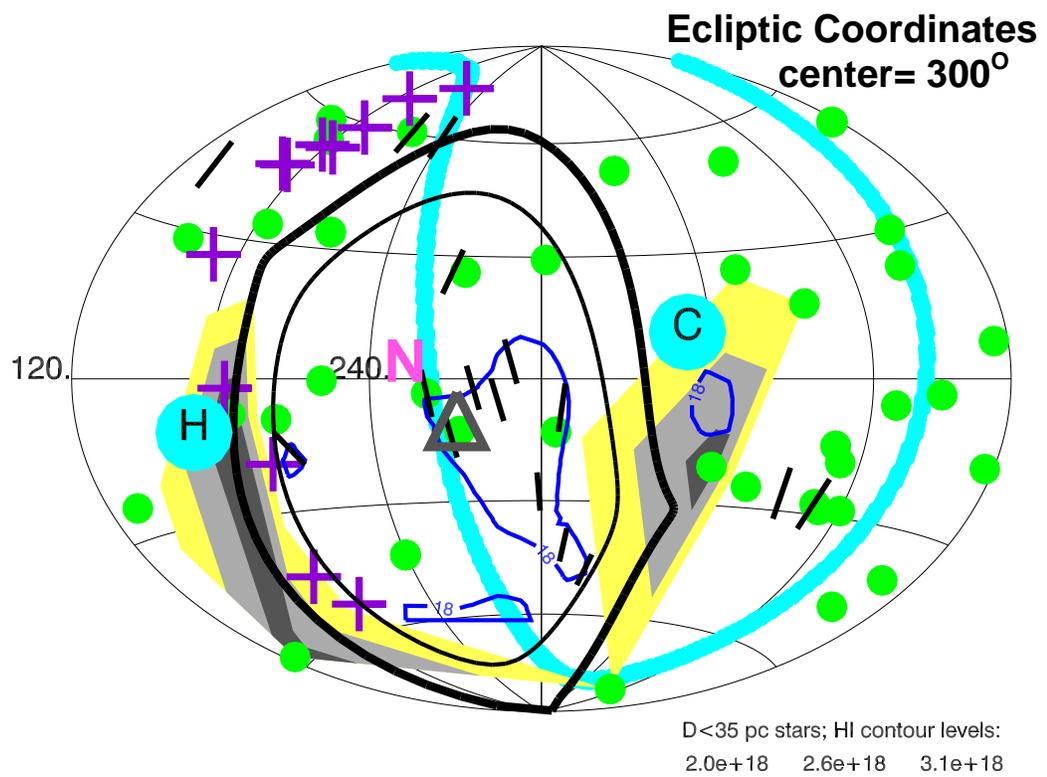}
\caption{Same as Fig. \ref{fig:localfluff1}, but for stars within 35 pc. \label{fig:localfluff2}}
\end{figure}

\begin{figure}[t]
\plotone{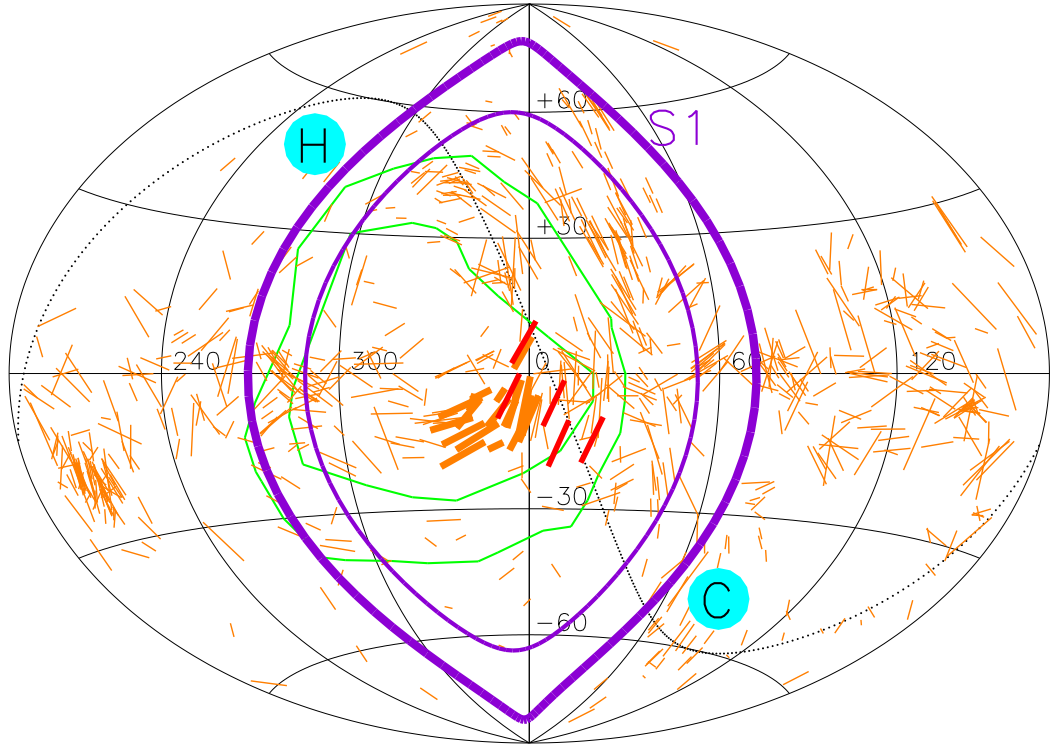}
\caption{ Polarization data for stars within 250 pc from
\citet[][orange bars]{Heiles:2000pol}, and polarizations of nearby
stars in the TPC that form the polarization maximum in the ecliptic
plane (red bars, \S \ref{sec:data}).  The figure is an Aitoff
projection with galactic coordinates.
The S1 subshell of the Loop I magnetic bubble is indicated (purple lines, 
Wolleben 2007), as is the ring attributed to the interaction of the
Loop I and Local Bubble \citep[green circles,][]{Egger:1995}.  The H
and C symbols and cyan-colored dots indicate the hot and cold poles of
the CMB dipole moment.  The near and far ISM yield the same position angles, to
within uncertainties, indicating that Loop I dominates the large-scale
magnetic field structure in the upwind direction.
\label{fig:loop} }
\end{figure}

\begin{figure}
\plotone{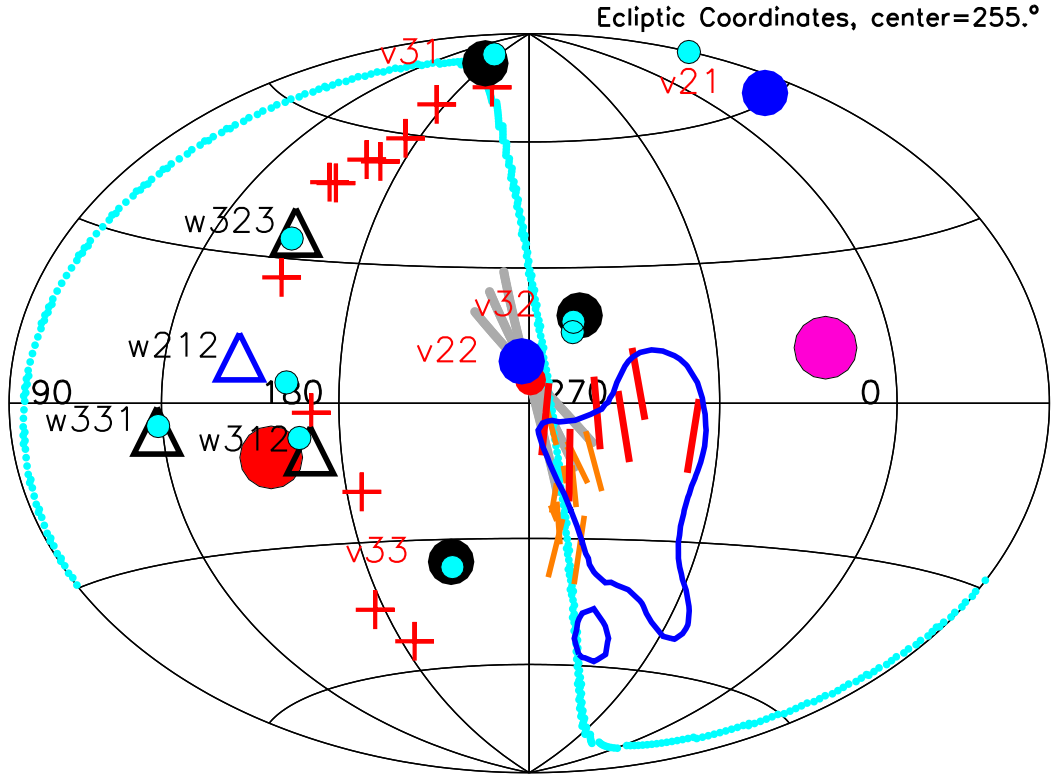}
\caption{ The positions of ILC123 multipole vectors $\hat{v}$ and area
vectors $\stackrel{\rightarrow}{w}$, that carry the phase information
about the CMB power for $\ell=$2 and $\ell=3$, are plotted in ecliptic
coordinates \citep[from][Table 1, ILC123 data]{Copietal:2007}.  The
quadrupole multipole vectors $v^{21}$ and $v^{22}$ (filled blue dots),
and area vector $w^{212}$ are the blue dots and triangle.  The
octopole multipole vectors $v^{31}$$v^{32}$, and $v^{33}$ and area
vectors, $w^{312}$, $w^{323}$, $w^{331}$, are plotted as black dots
and triangles, respectively.  The other symbols are the same as in
Fig. \ref{fig:ilc3}.  The $v^{22}$ multipole vector coincides with the
heliosphere nose direction.  The multipoles for the ILC1 map (from
Table 1, Copi et al., 2007) are plotted as small cyan dots.  The four
area vectors (normals to planes formed by multipole pairs) and the hot
pole of the CMB dipole are all located towards the sidewind band
defined by the 3 kHz emissions.  The 3 kHz emissions are formed where
outward propagating global merged interaction regions interact with
the magnetically-shaped heliopause \citep{MitchellCairnsetal:2004}.
The coincidence of the area vectors with the 3 kHz emission band
suggests that the ecliptic coordinate system is felt by the low-$\ell$
moments of the CMB because of the shaping of the heliosphere
by the interstellar magnetic field.  The plot is in ecliptic
coordinates and is centered on the heliosphere nose
at $\lambda \sim  255^\circ$. }\label{fig:copi}
\end{figure}

\begin{figure}[t!]
\includegraphics[angle=90,scale=0.8]{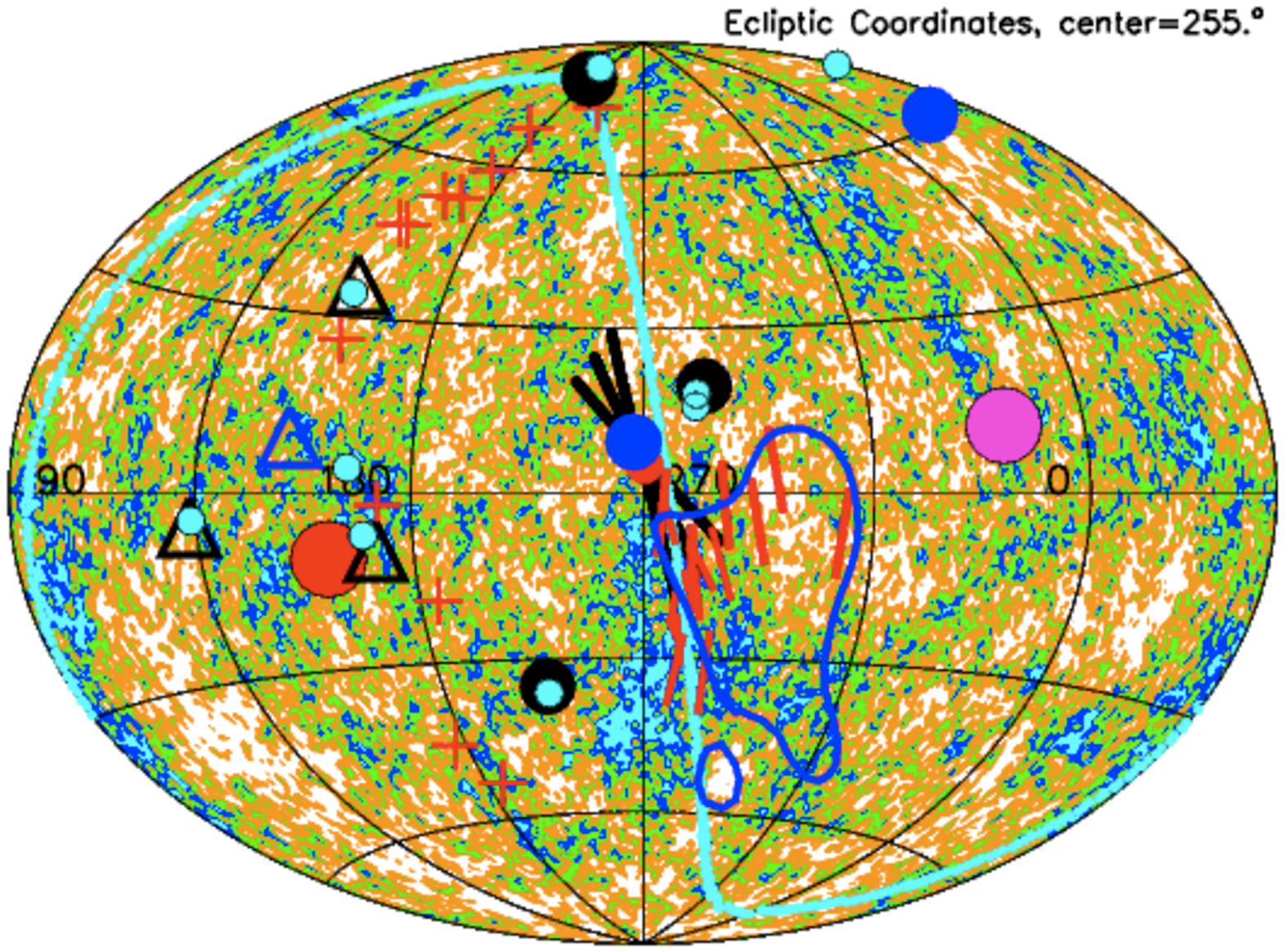}
\caption{ \small The WMAP Internal Linear Combination for Years 1-3
is displayed in the ecliptic coordinate system, and shifted by 105\deeg\
so that the LIC upwind direction is at the plot center.  Ecliptic
longitude is labeled, and increases to the right.  
The large red (pink) dot shows the maximum (minimum)
temperature of the CMB dipole moment (Table \ref{tab:direct}).
The thick cyan line shows
the plane that is equidistant between the CMB dipole temperature maximum
and minimum.  The small partially hidden red dot shows the observed upwind direction of
interstellar gas and large dust grains flowing into the heliosphere.
The three thick black lines show the offset angle,
and uncertainties on that angle, between the inflow directions of
interstellar \HeI\ and \HI.  The red bars show the stellar
polarizations from Fig. \ref{fig:copi}. 
The red crosses show the 
positions of the 3 kHz emission events
(Appendix \ref{app:khz}).  
The CMB data temperature scale is: $dT$: $< -0.119$ (cyan), --0.082 to
--0.055 (blue), --0.055 to 0.0 (green), 0 to 0.055 (orange), $> 0.055$
(white), where the temperature $dT$ is in mK.  
} \label{fig:ilc3}
\end{figure}

\begin{figure}
\plotone{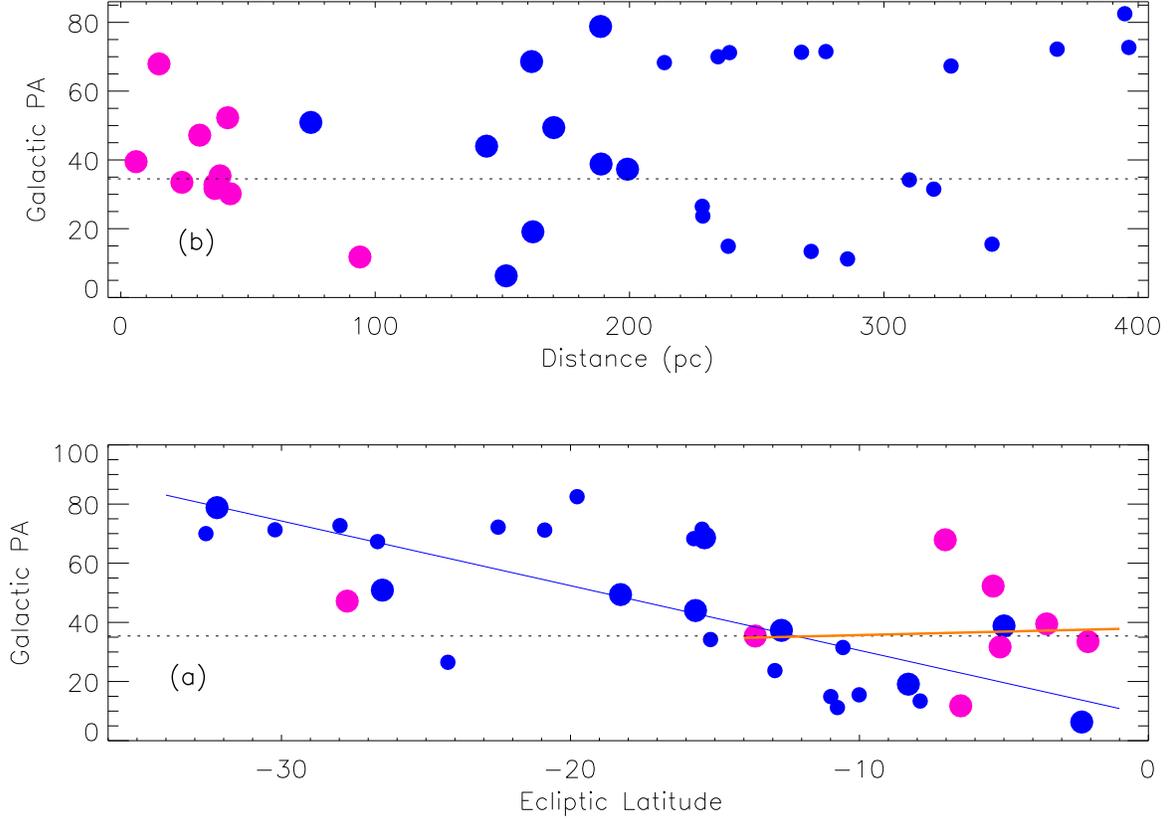}
\caption{ Position angles in galactic coordinates (\PAgal) are plotted
against the ecliptic latitude (a) and star distance (b) for stars at
negative ecliptic latitudes in the heliosphere nose region (Appendix
\ref{app:pa}).  Large (small) symbols indicate stars with distances $D
<200$ pc ($D=200-400$ pc).  Blue (pink) symbols indicate stars from
the HPC (TPC).  The \PAgal\ values for nearby stars, $D<200$ pc, is
relatively constant close to the ecliptic plane, $-15^\circ < \beta <
0^\circ$, with a value close to the \PAgal\ of the \HI-\HeI\ offset
angle (dotted lines).  The \PAgal\ values of distant stars show no
dependence on distance, but a strong dependence on ecliptic latitude
that traces the curvature of the magnetic field in the Loop I
superbubble in this region.  The blue line in (a) shows a first order
polynomial fit to the latitude dependence of \PAgal\ for HPC stars in
this region, $D < 400$ pc and $-35 < \beta < 0^\circ$.  The orange
line indicates a first order polynomial fit to the latitude dependence
of \PAgal\ for nearby TPC and HPC stars, $D < 200$ pc and $-15 < \beta
< 0^\circ$.  The dotted lines give the galactic position angle of the
\HI-\HeI\ offset angle.  Region selection boundaries and the
polynomial coefficients are given in Appendix \ref{app:pa}.
}\label{fig:poly}
\end{figure}

Voyagers 1 and 2 observed the 3 kHz emissions during the years 1992--1994.
The locations of the 3 kHz emissions are determined by triangulation
with Voyagers 1 and 2, the lag time for the solar flare material to
reach the outer heliosphere, and spacecraft pointing considerations
that included forbidden regions in the upwind directions.  The
result is that each emission event has two possible locations.  For
simplicity, the location closest to the heliosphere nose was selected as
the primary location by \citet{KurthGurnett:2003}, with the remaining
solution selected as the ambiguous (or alternate) location.  However, the true source
locations may be the reverse of this assumption, or even a combination 
of the primary and alternate locations.  The
primary solutions yield emissions that are aligned parallel to the
galactic plane in the upwind hemisphere, with an average galactic
latitude of $14.7^\circ \pm 7.3^\circ$, compared to the latitude of
the heliosphere nose of \glat=15.9\deeg.  If the primary solutions for
these events are assumed to form along a straight line in the galactic
coordinate system, then the best fit to that line is $b = 14.6 + 1.6
\times 10^{-3} * \ell $.  
The set of 3 kHz emission
locations that are best aligned with the $\lambda=181^\circ$
meridian consists of a combination of alternate and primary locations.
The emission event locations depicted in Figures
\ref{fig:localfluff1}, \ref{fig:localfluff2}, \ref{fig:ilc3}, and
\ref{fig:copi} consist of the alternate locations, 
except for the low latitude events of August 1993 at 2.95 kHz (6')
and November 1993 at 3.2 kHz (7) for which the primary locations are used.

Most of the theoretical attention has been
focused on the upwind solutions
\citep[e.g.][]{MitchellCairnsetal:2004}.  The alternate locations form
a pattern that samples a side-wind direction, with an average
$ecliptic$ longitude of \el=$181.4 \pm 19.0$.  This sidewind direction
is $\sim 73^\circ$ away from the upstream meridian, \el=254.7\deeg\
(Fig. \ref{fig:copi}).  Because of the coincidence of these sidewind
solutions with the CMB quadrupole area vector directions (\S
\ref{sec:cmb}, Fig. \ref{fig:copi}), the sidewind locations now
require theoretical attention.


\begin{deluxetable}{lccc}
\tablecolumns{4}
\tablecaption{Position Angles\label{tab:pa}}
\tablewidth{0pt}
\tablehead{
\colhead{Item} & \colhead{\PAgal} & \colhead{\PAecl} & \colhead{Notes}\\
\colhead{}&\colhead{}&\colhead{}&\colhead{} } 
\startdata
Tinbergen (1982)\tablenotemark{A} & $35 \pm 4$  &  $-25 \pm 4$ & Five upwind stars \\
&  &  & within 40 pc and \P5$\ge 17.5$ \\
\citet{Heiles:2000pol}\tablenotemark{B} & $41 \pm 20$ & $-20 \pm 20$  & Stars near polarization  \\
&  &  & maximum and at 140--240 pc \\
CMB dipole moment \tablenotemark{C,D} & $50^\circ \pm 1^\circ$  & $-11^\circ \pm 1^\circ$ &  5.0\deeg\ from \HeI\ upwind \\
\HI - \HeI\ offset\tablenotemark{D} & $27^\circ \pm 14^\circ$  & $-33^\circ \pm 14^\circ $ & \\
\enddata
\tablenotetext{A}{Position angle averages are calculated for
stars in the interval 
$260^\circ < \lambda < 305^\circ$, 
$-25^\circ < \beta < 10^\circ$}
\tablenotetext{B}{Position angle averages are calculated for
stars in the interval 
$260^\circ < \lambda < 305^\circ$, 
$-25^\circ < \beta < 10^\circ$ and
$ 335 < \ell < 360^\circ$, 
$-18^\circ  < b < 10^\circ $.}

\tablenotetext{C}{Position angles represent the position angle
of the plane that is equidistant between the two
poles of the CMB dipole moment, 
for the location \el$\sim 260.3^\circ$, \eb$\sim +6.6^\circ$, which is the
closest point of the this plane to the \HeI\ upstream direction.  Th
is plane separates the hot and cold hemispheres of the CMB Doppler dipole moment.}

\tablenotetext{D}{Position angles are calculated using the IDL routine posang.pro,
which gives the position angle of point 2 with respect to a great circle
meridian passing through point 1.}
\end{deluxetable}


\begin{deluxetable}{lccc}
\tablecolumns{4}
\tablecaption{Summary of Directions \label{tab:direct}}
\tablewidth{0pt}
\tablehead{
\colhead{Direction }&\colhead{\glong,\glat}&\colhead{\el,\eb }&\colhead{Ref.}}
\startdata
Tinbergen Polarization Max.& $\sim 19.2$\deeg, $\sim -21.2$\deeg & 294\deeg$\pm$10\deeg,  $0 \pm 20^\circ$& \S \ref{sec:data} \\
\HeI\ upwind (LIC) &  $3.5^\circ$ +15.2\deeg\  & $255.4^\circ \pm 0.5^\circ$, $5.1^\circ \pm  0.5^\circ$   & 1 \\
\HI\ upwind & 5.8\deeg, +19.3\deeg  & $252.9^\circ \pm 0.5^\circ$, $9.0 \pm 0.5^\circ$  & 2 \\
CMB dipole hot pole & 263.9\deeg, 48.3\deeg & 171.6, --11.1  & 3  \\
CMB dipole plane to upstream\tablenotemark{A} & $6.9^\circ \pm 0.5^\circ$, $11.4^\circ \pm 0.1^\circ$  & $260.4^\circ \pm 0.5^\circ$, $6.1^\circ \pm 0.5^\circ$  & 3  \\
ISM Dust upwind & 263.9\deeg, 48.3\deeg & $259^\circ \pm 10^\circ$, $8^\circ \pm 5^\circ$ & 4 \\
LIC LSR upstream & 346\deeg, --1\deeg & 260\deeg, --18\deeg\  & 5 \\ 
CLIC LSR upstream & 0\deeg, --5\deeg & 271\deeg, --8\deeg\  & 5 \\ 
\enddata
\tablerefs{
1.  \citet[][and private communication]{Witte:2004}.
2.  \citet{Lallementetal:2005}.
3.  This paper and \citet{Hinshawetal:2007}[Hinshaw 2007 reference]
4.  \citet{Frischetal:1999}.
5.  \citet{FGW:2002}, for solar apex motion from \citet{DehnenBinney:1998}.
}
\tablenotetext{A}{This plane is equidistant between the hot and cold
poles of the CMB dipole moment.}
\end{deluxetable}

\begin{deluxetable}{lcccc}
\tablecolumns{4}
\tablecaption{Stars Dominating Ecliptic Polarization\label{tab:star}}
\tablewidth{0pt}
\tablehead{
\colhead{HD}&\colhead{Dist.} & &\colhead{\P5 } & \colhead{\PAecl} }
\startdata
155885 (36 Oph) &  6 && 17.5 &  --20\deeg \\
161892 & 39  && 25.5 &   --25\deeg  \\
169916 ($\lambda$ Sgr) &  24 && 17.5 &  --27\deeg  \\
177716 ($\tau$ Sgr) & 37 & & 21.2 &   --29\deeg  \\
181577 ($\rho^1$ Sgr) & 37 & & 21.9 &  --25\deeg \\
\enddata
\end{deluxetable}


\newpage
\newpage

\newpage

\newpage

\newpage

\newpage

\newpage

\newpage
\clearpage

\end{document}